\documentclass[useAMS,usenatbib]{mn2e}
\usepackage{mn2e-breakabs}
\usepackage{graphicx}
\usepackage{subfigure}
\usepackage{times}
\newcommand{\Hunit}{\,{\rm km}\,{\rm s}^{-1}\,{\rm Mpc}^{-1}}

\def\la{\mathrel{\mathpalette\fun <}}
\def\ga{\mathrel{\mathpalette\fun >}}
\def\fun#1#2{\lower3.6pt\vbox{\baselineskip0pt\lineskip.9pt
        \ialign{$\mathsurround=0pt#1\hfill##\hfil$\crcr#2\crcr\sim\crcr}}}

\def\bfs{\mbox{\bf s}}

\newcommand{\be}{\begin{equation}}
\newcommand{\ee}{\end{equation}}
\newcommand{\ba}{\begin{eqnarray}}
\newcommand{\ea}{\end{eqnarray}}
\newcommand{\simgt}{\,\hbox{\lower0.6ex\hbox{$\sim$}\llap{\raise0.6ex\hbox{$>$}}}\,}
\newcommand{\simlt}{\,\hbox{\lower0.6ex\hbox{$\sim$}\llap{\raise0.6ex\hbox{$<$}}}\,}

\begin{document}

\title[SDSS DR7 LRG 2D Correlation Function]
{Measurements of $H(z)$ and $D_A(z)$ from
the Two-Dimensional Two-Point Correlation Function of 
Sloan Digital Sky Survey Luminous Red Galaxies}

\author[Chuang \& Wang]{
  \parbox{\textwidth}{
 Chia-Hsun Chuang\thanks{E-mail: chuang@nhn.ou.edu}
 and Yun Wang
  }
  \vspace*{4pt} \\
  Homer L. Dodge Department of Physics \& Astronomy, Univ. of Oklahoma,
                 440 W Brooks St., Norman, OK 73019, U.S.A.\\
}

\date{\today} 

\maketitle

\begin{abstract}

We present a method for measuring the Hubble parameter, $H(z)$, and 
angular diameter distance, $D_A(z)$, from the two-dimensional 
two-point correlation function, and validate it using LasDamas mock galaxy catalogs.
Applying our method to the sample of luminous red galaxies (LRGs) 
from the Sloan Digital Sky Survey (SDSS) Data Release 7 (DR7), we measure
$H(z=0.35)\equiv H(0.35)=82.1_{-4.9}^{+4.8}$ $\Hunit$, 
$D_A(z=0.35)\equiv D_A(0.35)=1048_{-58}^{+60}$ Mpc
without assuming a dark energy model or a flat Universe.
We find that the derived measurements of $H(0.35)\,r_s(z_d)/c$ and 
$D_A(0.35)/r_s(z_d)$ (where $r_s(z_d)$ is the sound horizon 
at the drag epoch) are nearly uncorrelated, have tighter constraints 
and are more robust with respect to possible systematic effects. 
Our galaxy clustering measurements of 
$\{H(0.35)\,r_s(z_d)/c, D_A(0.35)/r_s(z_d)\}=\{0.0434\pm 0.0018 ,6.60\pm0.26\}$ 
(with the correlation coefficient $r = 0.0604$) can be used to combine with 
cosmic microwave background and any other cosmological
data sets to constrain dark energy. Our results represent the first 
measurements of $H(z)$ and $D_A(z)$ (or $H(z)\,r_s(z_d)/c$ and $D_A(0.35)/r_s(z_d)$) 
from galaxy clustering data.
Our work has significant implications for future surveys 
in
establishing the feasibility of measuring both $H(z)$ and $D_A(z)$
from galaxy clustering data.

\end{abstract}


\section{Introduction}
The cosmic large-scale structure from galaxy redshift surveys provides a 
powerful probe of dark energy and the cosmological model
that is highly complementary to the cosmic microwave 
background (CMB) \citep{Bennett03}, supernovae (SNe) 
\citep{Riess:1998cb,Perlmutter:1998np}, and weak lensing
\citep{Wittman00,bacon00,kaiser00,vw00}. The scope of galaxy redshift 
surveys has dramatically increased in the last decade. The PSCz 
surveyed $\sim 15,000$ galaxies using the Infrared Astronomical Satellite (IRAS)  
\citep{Saunders:2000af}, the 2dF Galaxy Redshift Survey (2dFGRS) 
obtained 221,414 galaxy redshifts \citep{Colless:2001gk,Colless:2003wz}, 
and the Sloan Digital Sky Survey (SDSS) has collected 
930,000 galaxy spectra in the Seventh Data Release (DR7) \citep{Abazajian:2008wr}.
The ongoing galaxy surveys will probe the Universe at higher redshifts;
WiggleZ is surveying 240,000 emission-line galaxies at $0.5<z<1$ over 
1000 square degrees \citep{Blake09}, and BOSS is surveying 1.5 million 
luminous red galaxies (LRGs) at $0.1<z<0.7$ over 10,000 square degrees \citep{Eisenstein:2011sa}.
The planned space mission Euclid will survey over 60 million
emission-line galaxies at $0.5<z<2$ over 20,000 square degrees \citep{Cimatti09,Wang10}.

Large-scale structure data from galaxy surveys can be analyzed using either 
the power spectrum or the correlation function. Although these two methods are simple 
Fourier transforms of one another, the analysis processes are quite 
different and the results cannot be converted using Fourier transform 
directly because of the finite size of the survey volume. 
The SDSS data have been analyzed using both the power spectrum 
method (see, e.g., \citealt{Tegmark:2003uf,Hutsi:2005qv,Padmanabhan:2006ku,Blake:2006kv,Percival:2007yw,Percival:2009xn,Reid:2009xm}), 
and the correlation function method (see, e.g., 
\citealt{Eisenstein:2005su,Okumura:2007br,Cabre:2008sz,Martinez:2008iu,Sanchez:2009jq,Kazin:2009cj,Chuang:2010dv}). 
While previous work has focused on the spherically averaged two-point correlation function (2PCF), or the radial projection
of the two-dimensional two point correlation function 
(2D 2PCF), we measure and analyze the full 2D 2PCF of SDSS LRGs in this study.

The power of galaxy clustering as a dark energy probe
lies in the fact that the Hubble parameter, $H(z)$, and the
angular diameter distance, $D_A(z)$, can in principle
be extracted simultaneously from data through the measurement
of the baryon acoustic oscillation (BAO) scale 
in the radial and transverse directions \citep{BG03,SE03,Wang06}. 
This has not been achieved in the previous work in the
analysis of real data. 
\cite{Okumura:2007br} concluded that SDSS DR3 LRG data were
not sufficient for measuring $H(z)$ and $D_A(z)$;
they derived constraints on cosmological parameters assuming that
dark energy is a cosmological constant. \cite{Cabre:2008sz} measured 
the linear redshift space distortion parameter $\beta$, galaxy bias, 
and $\sigma_8$ from SDSS DR6 LRGs. \cite{Gaztanaga:2008xz} 
obtained a measurement of $H(z)$ by measuring the peak of 
the 2PCF along the line of sight. However, \cite{Kazin:2010nd}
showed that the amplitude of the line-of-sight peak 
is consistent with sample variance. 

In our previous paper, \cite{Chuang:2010dv}, we presented the method to 
obtain dark energy and cosmological model constraints from the spherically-averaged 2PCF,
without assuming a dark energy model or a flat Universe. We  
demonstrated the feasibility of extracting $H(z)$ and $D_A(z)$
by scaling the spherically-averaged 2PCF (which leads to highly
correlated measurements). In this paper, we obtain robust measurements of 
$H(z)$ and $D_A(z)$ through scaling, using the 2D correlation function measured
from a sample of SDSS DR7 LRGs \citep{Eisenstein:2001cq}.
This sample is homogeneous and has the largest effective survey volume 
to date for studying the quasi-linear regime \citep{Eisenstein:2005su}. 
In Section \ref{sec:data}, we introduce the galaxy sample used 
in our study. In Section \ref{sec:method}, we describe the details of our 
method. In Section \ref{sec:results}, we present our results. In Sectioin \ref{sec:test},
we apply some systematic tests to our measurements.
We summarize and conclude in Sec.~\ref{sec:conclusion}.

\section{Data} \label{sec:data}

The SDSS has observed one-quarter of the
entire sky and performed a redshift survey of galaxies, quasars and
stars in five passbands $u,g,r,i,$ and $z$ with a 2.5m telescope
\citep{Fukugita:1996qt,Gunn:1998vh,Gunn:2006tw}. 
We use the public catalog, the NYU Value-Added Galaxy Catalog
(VAGC) \citep{Blanton:2004aa}, derived from the
SDSS II final public data release, Data Release 7 (DR7)
\citep{Abazajian:2008wr}.
We select our LRG sample from the NYU VAGC with 
the flag $primTarget$ bit mask set to $32$. K-corrections
have been applied to the galaxies with a
fiducial model ($\Lambda$CDM with $\Omega_m=0.3$ and $h=1$), and
the selected galaxies are required to have rest-frame $g$-band absolute
magnitudes $-23.2<M_g<-21.2$ \citep{Blanton:2006kt}. The same 
selection criteria were used in previous papers
\citep{Zehavi:2004zn,Eisenstein:2005su,Okumura:2007br,Kazin:2009cj}. 
The sample we use is referred to as ``DR7full'' in \cite{Kazin:2009cj}.
Our sample includes 87000 LRGs in the redshift range 0.16-0.44.
\footnote{We have identified a bug while computing the weighting of each galaxy in the first draft of this paper.
The bug was that we computed the weights of random data
with the number density of random data instead of observed data. This would introduce a bias 
when the number density is not homogeneous.} 

Spectra cannot be obtained for objects closer than 55 arcsec
within a single spectroscopic tile due to the finite size of the
fibers. To correct for these ``collisions'', the redshift of an object
that failed to be measured would be assigned to be the same as the
nearest successfully observed one. Both fiber 
collision corrections and K-corrections have been made in NYU-VAGC 
\citep{Blanton:2004aa}. The collision corrections applied here are 
different from what has been suggested in \cite{Zehavi:2004zn}. 
However, the effect should be small since we are using relatively large 
scale which are less affected by the collision corrections.

We construct the radial selection function as a cubic spline fit 
to the observed number density histogram with the width $\Delta
z=0.01$. The NYU-VAGC provides the description of the
geometry and completeness of the survey in terms of spherical
polygons. We adopt it as the angular selection function of our
sample. We drop the regions with completeness
below $60\%$ to avoid unobserved plates \citep{Zehavi:2004zn}. The 
Southern Galactic Cap region is also dropped.

\section{Methodology} 
\label{sec:method}

In this section, we describe the measurement of the correlation function
from the observational data, construction of the theoretical prediction, 
and the likelihood analysis that leads to constraints on dark energy and 
cosmological parameters.

\subsection{Measuring the Two-Dimensional Two-Point Correlation Function}

We convert the measured redshifts of galaxies to comoving distances 
by assuming a fiducial model, $\Lambda$CDM with $\Omega_m=0.25$. 
We use the two-point correlation function estimator given by 
\cite{Landy:1993yu}:
\begin{equation}
\xi(\sigma,\pi) = \frac{DD(\sigma,\pi)-2DR(\sigma,\pi)+RR(\sigma,\pi)}{RR(\sigma,\pi)},
\end{equation}
where $\pi$ is the separation along the light of sight (LOS), $\sigma$ 
is the separation in the plane of the sky, DD, DR, and RR represent the normalized 
data-data,
data-random, and random-random pair counts respectively in a
distance range. The LOS is defined as the direction from the observer to the 
center of a pair. The bin size we use in this study is
$10 \, h^{-1}$Mpc$\times 10 \,h^{-1}$Mpc. 
The Landy and Szalay estimator has minimal variance for a Poisson
process. Random data are generated with the same radial
and angular selection functions as the real data. One can reduce the shot noise due
to random data by increasing the number of random data. The number
of random data we use is 10 times that of the real data. While
calculating the pair counts, we assign to each data point a radial
weight of $1/[1+n(z)\cdot P_w]$, where $n(z)$ is the radial
selection function and $P_w = 4\cdot 10^4$ $h^{-3}$Mpc$^3$ 
\citep{Eisenstein:2005su}. 
We use the same $P_w$ as Eisenstein et al. (2005) in which they used the sample of the SDSS DR3. Although the data release versions are different, the properties of the galaxy sample should be basically the same. We find that the error bars estimated from LasDamas mock catalogs could be improved by 10\% while using the weighting (compared to the error bars obtained without using the weighting). We expect that the results should not be sensitive to the $P_w$ used.

\subsection{Theoretical Two-Dimensional Two-Point Correlation Function}
We compute the linear power spectra at $z=0.35$ by using CAMB
\citep{Lewis:1999bs}. To include the effect of non-linear structure
formation on the BAOs, we first calculate the dewiggled power spectrum 
\begin{equation} \label{eq:dewiggle}
P_{dw}(k)=P_{lin}(k)\exp\left(-\frac{k^2}{2k_\star^2}\right)
+P_{nw}(k)\left[1-\exp\left(-\frac{k^2}{2k_\star^2}\right)\right],
\end{equation}
where $P_{lin}(k)$ is
the linear matter power spectrum, $P_{nw}(k)$ is the no-wiggle or pure
CDM power spectrum calculated using Eq.(29) from \cite{Eisenstein:1997ik}, 
and $k_{\star}$ is marginalized over\footnote{Although $k_{\star}$ can be computed by 
renormalization perturbation theory \citep{Crocce:2005xz,Matsubara:2007wj}, 
doing so requires knowing the amplitude of the power spectrum, which is 
also marginalized over in this study.} with a flat prior over the range of
0.09 to 0.13 $h$Mpc$^{-1}$.

We then use the software package \emph{halofit} \citep{Smith:2002dz} to compute the 
non-linear matter power spectrum:
\begin{eqnarray} \label{eq:halofit}
r_{halofit}(k) &\equiv& \frac{P_{halofit,nw}(k)}{P_{nw}(k)} \\
P_{nl}(k)&=&P_{dw}(k)r_{halofit}(k),
\end{eqnarray}
where $P_{halofit,nw}(k)$ is the power spectrum obtained by applying halofit 
to the no-wiggle power spectrum, and $P_{nl}(k)$ is the non-linear power spectrum. 
We compute the theoretical real space two-point correlation 
function, $\xi(r)$, by Fourier transforming the non-linear power spectrum
$P_{nl}(k)$. 

In the linear regime (i.e., large scales) and adopting the small-angle approximation
(which is valid on scales of interest), the 2D correlation function in the redshift space can 
be written as 
\citep{Kaiser:1987qv,hamilton1992}
\begin{equation}
 \xi^{\star}(\sigma,\pi)=\xi_0(s)P_0(\mu)+\xi_2(s)P_2(\mu)+\xi_4(s)P_4(\mu),
\end{equation}
where $s=\sqrt{\sigma^2+\pi^2}$, 
$\mu$ is the cosine of the angle between $\bfs=(\sigma,\pi)$ and the LOS, and 
$P_l$ are Legendre polynomials. The multipoles of $\xi$ are defined as
\begin{eqnarray}
 \xi_0(r)&=&\left(1+\frac{2\beta}{3}+\frac{\beta^2}{5}\right)\xi(r),\\
 \xi_2(r)&=&\left(\frac{4\beta}{3}+\frac{4\beta^2}{7}\right)[\xi(r)-\bar{\xi}(r)],\\
 \xi_4(r)&=&\frac{8\beta^2}{35}\left[\xi(r)+\frac{5}{2}\bar{\xi}(r)
 -\frac{7}{2}\overline{\overline{\xi}}(r)\right],
\end{eqnarray}
where $\beta$ is the redshift space distortion parameter and
\begin{eqnarray}
 \bar{\xi}(r)&=&\frac{3}{r^3}\int_0^r\xi(r')r'^2dr',\\
 \overline{\overline{\xi}}(r)&=&\frac{5}{r^5}\int_0^r\xi(r')r'^4dr'.
\end{eqnarray}
Next, we convolve the 2D correlation function with the distribution function of 
random pairwise velocities, $f(v)$, to obtain the final model $\xi(\sigma,\pi)$ 
\citep{peebles1980}
\begin{equation}
 \xi(\sigma,\pi)=\int_{-\infty}^\infty \xi^\star\left(\sigma,\pi-\frac{v}{H(z)a(z)}
 \right)\,f(v)dv,
\end{equation}
where the random motions are represented by an exponential form 
\citep{ratcliffe1998,Landy:2002xg}
\begin{equation}
 f(v)=\frac{1}{\sigma_v\sqrt{2}}\exp\left(-\frac{\sqrt{2}|v|}{\sigma_v}\right),
\end{equation}
where $\sigma_v$ is the pairwise peculiar velocity dispersion.

The parameter set we use to compute the theoretical
correlation function is 
$\{H(z), D_A(z), \beta, \Omega_mh^2, \Omega_bh^2, n_s, \sigma_v, k_\star\}$, where
$\Omega_m$ and $\Omega_b$ are the density fractions of matter and
baryons, $n_s$ is the powerlaw index of the primordial matter power spectrum, 
and $h$ is the dimensionless Hubble
constant ($H_0=100h$ km s$^{-1}$Mpc$^{-1}$). 
We set $h=0.7$ while calculating the non-linear power spectra. On the scales 
we use for comparison with data, the theoretical correlation 
function only depends on cosmic curvature and dark energy through
parameters $H(z)$ and $D_A(z)$, assuming that dark energy perturbations
are unimportant (valid in simplest dark energy models).
Thus we are able to extract constraints from data that are independent of a dark energy 
model and cosmic curvature. 

\begin{figure*}
\begin{center}
 \subfigure[2D 2PCF from SDSS LRGfull sample]{\label{fig:sdssfull_twod}\includegraphics[width=1 \columnwidth,clip,angle=0]{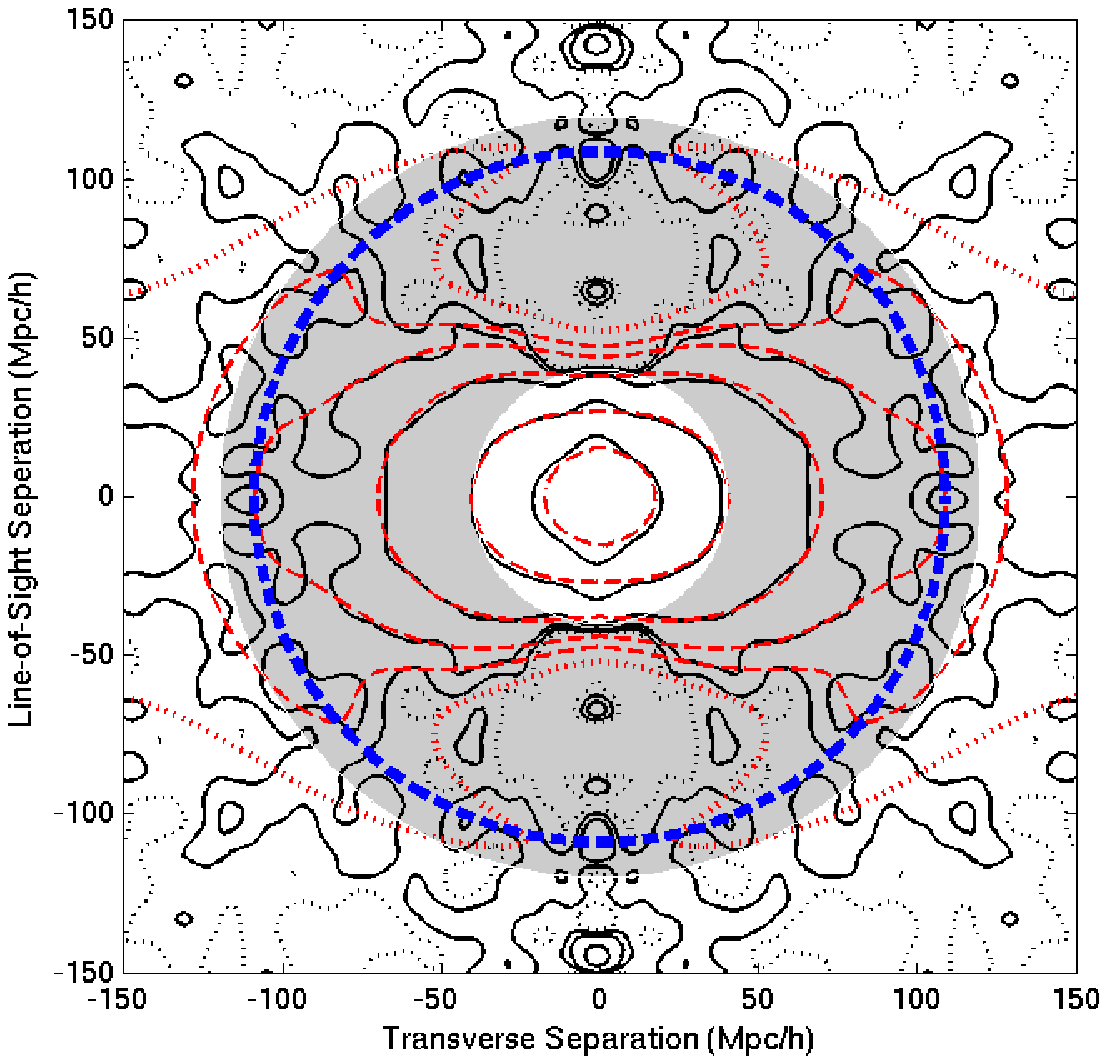}}
 \subfigure[2D 2PCF from single mock catalog]{\label{fig:lasdamas_twod-single}\includegraphics[width=1 \columnwidth,clip,angle=0]{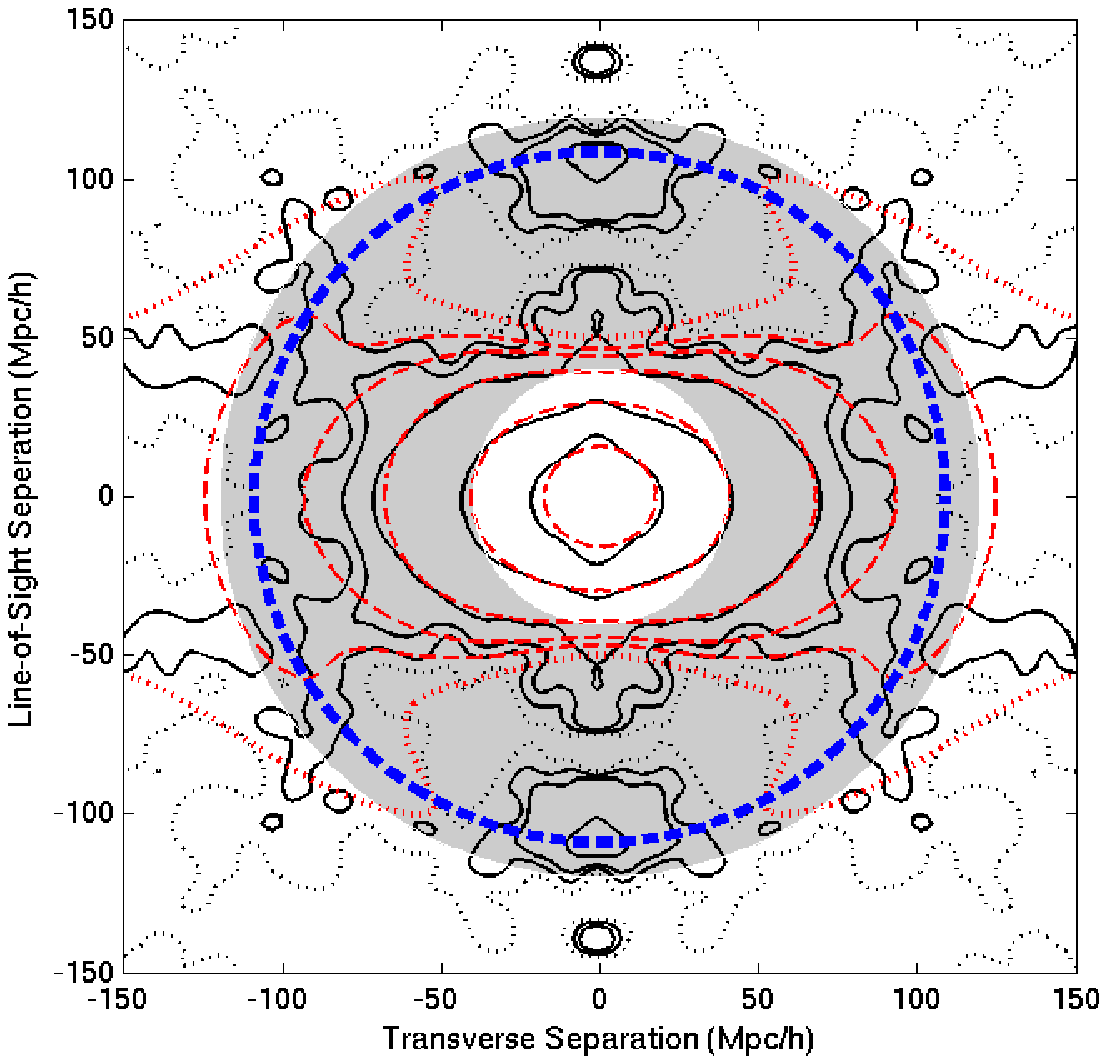}}
\end{center}
\caption{\subref{fig:sdssfull_twod} The two-dimensional two-point correlation function (2D 2PCF)
measured from SDSS DR7 LRGs in a redshift range  $0.16 < z < 0.44$ (solid black
contours), compared to a theoretical correlation function with 
parameters close to the best fit values in the likelihood analysis
(dashed red contours). The theoretical model has
$H(z=0.35)=81.8\Hunit$, $D_A(z=0.35)=1042 \,{\rm Mpc}$, $\beta=0.35$, $\Omega_mh^2=0.117$, 
$\Omega_bh^2=0.022$, $n_s=0.96$, $\sigma_v=300 {\rm km\ s^{-1}}$, and $k_{\star}=0.11$. 
\subref{fig:lasdamas_twod-single} The 2D 2PCF measured from a single mock catalog, compared to a theoretical model with the input parameters 
of the LasDamas simulations and $\{\beta, \sigma_v, k_\star\}$ are set to $\{0.35, 300$km\ s$^{-1}, 0.11h$Mpc$^{-1}\}$ (dashed red contours).
In both figures, the shaded disk indicates the scale range considered ($s=40-120\ h^{-1}$Mpc ) in this study.
The thick dashed blue circle denotes the baryon acoustic oscillation scale. 
The observed 2D 2PCF has been smoothed by a Gaussian filter with rms variance 
of $2 h^{-1}$Mpc for illustration in these figures only; smoothing is not used in our likelihood 
analysis. The contour levels are $\xi=0.5, 0.1, 0.025, 0.01, 0.005, 0$. 
The $\xi=0$ contours are denoted with dotted lines for clarity.
}
\label{fig:single_twod}
\end{figure*}

Fig.\ref{fig:sdssfull_twod} shows the 2D 2PCF measured from SDSS LRGs 
compared with a theoretical model. The measured 2D 2PCF of the SDSS LRGs has been 
smoothed by a Gaussian filter with rms variance of $2 h^{-1}$Mpc to
illustrate the comparison of data with model in this
figure, as the unsmoothed data are very noisy. Smoothing is {\it not} used 
in our likelihood analysis to avoid possibly introducing systematic biases.
Fig. \ref{fig:lasdamas_twod-single} shows the 2D 2PCF measured from a single
LasDamas SDSS LRG mock catalog for comparison. The similarity between the
data and the mock in the range of scales we used (indicated by the shaded
disk) is apparent.

\begin{figure}
\centering
\includegraphics[width=1 \columnwidth,clip,angle=0]{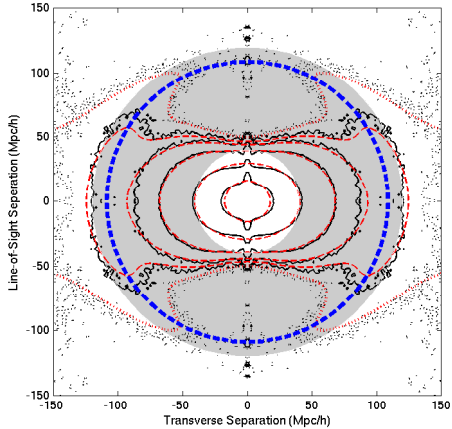}
\caption{The average two-dimensional two-point correlation
function (2D 2PCF) measured from 160 LasDamas SDSS LRGfull mock catalogs (solid black
contours), compared to a theoretical model with the input parameters 
of the LasDamas simulations and $\{\beta, \sigma_v, k_\star\}$ are set to $\{0.35, 300$km\ s$^{-1}, 0.11h$Mpc$^{-1}\}$ (dashed red contours).
The gray area is the scale range considered ($s=40-120\ h^{-1}$Mpc ) in this study.
The thick dashed blue circle denotes the baryon acoustic oscillation scale.
The contour levels are apparent in the 2D 2PCF measured from mock catalogs, 
even though no smoothing is used.
The contour levels are $\xi=0.5, 0.1, 0.025, 0.01, 0.005, 0$.
The $\xi=0$ contours are denoted with dotted lines for clarity.}
\label{fig:lasdamas_twod}
\end{figure}

Fig.\ref{fig:lasdamas_twod} shows the averaged 2D 2PCF 
measured from the LasDamas mock catalogs compared with a theoretical model.
The contour levels are apparent in the measured 2D 2PCF even though no smoothing
is used; this is due to the reduction of shot noise achieved by averaging over
160 mock catalogs.
Clearly, our 2D theoretical model
provides a reasonable fit to data on intermediate (and quasi-linear)
scales. The deviations on smaller scales may be due to the simplicity
of the peculiar velocity model we have used. 
We do not use the smaller scales ($s<40\ h^{-1}$Mpc),
where the scale dependence of redshift distortion and galaxy bias are not
negligible and cannot be accurately determined at present. 
According to Fig. 5 in \cite{Eisenstein:2005su} and Fig. 4 in \cite{Blake:2011wn}, 
these effects are negligible at $s>40\ h^{-1}$Mpc.
On large scales, data become
very noisy as sample variance dominates. For these reasons, we will only 
use the scale range of $s=40-120\,h^{-1}$Mpc in our analysis.
We do not consider wide-angle effects, since they have been shown to be small 
on the length scales of interest here \citep{Samushia:2011cs}.
\cite{Samushia:2011cs} showed that the corrections (i.e. nonlinear effect and wide-angle effect) to the Kaiser formula are small comparing to the statistical errors on the measurement of the correlation function from SDSS DR7 LRG for the scale range interested (s=40-120 Mpc/h). In this study, we include the largest correction, dewiggling (nonlinear-BAO), and the nonlinear effects at small scales.
Since including a larger range of scales gives more stringent 
constraints, our choice of $s=40-120\,h^{-1}$Mpc represents
a conservative cut in data to reduce contamination by systematic
uncertainties.

\subsection{Covariance Matrix} \label{sec:covar}

We use the mock catalogs from the LasDamas 
simulations\footnote{http://lss.phy.vanderbilt.edu/lasdamas/} 
(McBride et al., in preparation) 
to estimate the covariance matrix of the observed correlation function. 
LasDamas provides mock catalogs matching SDSS main galaxy and LRG samples.
We use the LRG mock catalogs from the LasDamas gamma release with the same cuts as
the SDSS LRG DR7full sample, $-23.2<M_g<-21.2$ and $0.16<z<0.44$.
We have diluted the mock catalogs to match the radial selection function 
of the observed data by randomly selecting the mock galaxies according to the 
number density of the data sample. We calculate the 2D correlation functions 
of the mock catalogs and construct the covariance matrix as
\begin{equation}
 C_{ij}=\frac{1}{N-1}\sum^N_{k=1}(\bar{\xi}_i-\xi_i^k)(\bar{\xi}_j-\xi_j^k),
\label{eq:covmat}
\end{equation}
where $N$ is the number of the mock catalogs, $\bar{\xi}_m$ is the
mean of the $m^{th}$ bin of the mock catalog correlation functions, and
$\xi_m^k$ is the value in the $m^{th}$ bin of the $k^{th}$ mock
catalog correlation function. Note that the covariance matrix constructed from the 
LasDamas mock catalogs is noisy because only 160 mock catalogs are available. 
Therefore, we smooth it before using it to calculate the likelihood.

\subsection{Likelihood}
The likelihood is taken to be proportional to $\exp(-\chi^2/2)$ \citep{press92}, 
with $\chi^2$ given by
\begin{equation} \label{eq:chi2}
 \chi^2\equiv\sum_{i,j=1}^{N_{bins}}\left[\xi_{th}(\bfs_i)-\xi_{obs}(\bfs_i)\right]
 C_{ij}^{-1}
 \left[\xi_{th}(\bfs_j)-\xi_{obs}(\bfs_j)\right]
\end{equation}
where $N_{bins}$ is the number of bins used, $\bfs_m=(\sigma_m,\pi_m)$, 
$\xi_{th}$ is the theoretical correlation function, and $\xi_{obs}$ 
is the observed correlation function. Note that $\xi_{th}(\bfs_i)$ depends
on $\{H(z), D_A(z), \beta, \Omega_mh^2, \Omega_bh^2, n_s, \sigma_v, k_\star\}$.

In principle, we should recalculate the observed correlation function while 
computing the $\chi^2$ for different models. However, since we don't consider 
the entire scale range of the correlation function (we only consider 
$s=40-120\ h^{-1}$Mpc in this study), we might include or exclude different 
data pairs for different models which would render $\chi^2$ values arbitrary. 
Therefore, instead of recalculating the observed correlation function, we 
apply the inverse operation to the theoretical correlation function
to move the parameter dependence from the data to the model,
thus preserving the number of galaxy pairs used in the likelihood analysis.

Let us define $T$ as the operator converting the 
measured correlation function from the fiducial model to another 
model, i.e.,
\begin{equation}
 \xi_{obs}(\bfs)=T(\xi_{obs}^{fid}(\bfs)),
\label{eq:T def}
\end{equation}
where $\xi^{fid}_{obs}(\bfs)$ is the observed correlation function 
assuming the
fiducial model. This allows us to rewrite $\chi^2$ as
\ba 
\label{eq:chi2_2}
\chi^2 &\equiv&\sum_{i,j=1}^{N_{bins}}
 \left\{T^{-1}\left[\xi_{th}(\bfs_i)\right]-\xi^{fid}_{obs}(\bfs_i)\right\}
 C_{fid,ij}^{-1} \cdot \nonumber\\
 & & \cdot \left\{T^{-1}\left[\xi_{th}(\bfs_j)\right]-\xi_{obs}^{fid}(\bfs_j)\right\},
\ea
where we have used Eqs.(\ref{eq:covmat}) and (\ref{eq:T def}).

To find the operator $T$, note that the fiducial model is only used
in converting redshifts into distances for the galaxies in our data
sample. In the analysis of galaxy clustering, we only need the
separation of a galaxy pair, and not the absolute distances to
the galaxies. For a thin redshift shell, 
the separation of a galaxy pair in the transverse direction is proportional to $D_A(z)\Delta\theta$ ($\Delta\theta$ is the angle separation of the galaxy pair) and the separation along the direction of the line of sight is proportional to $\Delta z /H(z)$ ($\Delta z$ is the redshift difference between the galaxy pair). Thus,
we can convert the separation of one pair of galaxies from the fiducial model to 
another model by performing the scaling
(see, e.g., \cite{SE03})
\begin{equation}
 (\sigma',\pi')=\left(\frac{D_A(z)}{D_A^{fid}(z)}\sigma,
 \frac{H^{fid}(z)}{H(z)}\pi\right).
\end{equation}
Therefore, we can convert the measured 2D correlation function from some model to 
the fiducial model as follows:
\begin{eqnarray}
 \xi^{fid}_{obs}(\sigma,\pi)&=&T^{-1}(\xi_{obs}(\sigma,\pi))\nonumber\\
 &=&\xi_{obs}\left(\frac{D_A(z)}{D_A^{fid}(z)}\sigma,
 \frac{H^{fid}(z)}{H(z)}\pi\right).\nonumber\\
\end{eqnarray}
This mapping defines the operator $T$.

We now apply the inverse operation to the theoretical correlation function:
\begin{equation} \label{eq:inverse_theory_2d}
 T^{-1}(\xi_{th}(\sigma,\pi))=\xi_{th}
 \left(\frac{D_A(z)}{D_A^{fid}(z)}\sigma,
 \frac{H^{fid}(z)}{H(z)}\pi\right).
\end{equation}
$\chi^2$ can be calculated by substituting eq.\ (\ref{eq:inverse_theory_2d}) 
into eq.\ (\ref{eq:chi2_2}).

\subsection{Markov Chain Monte-Carlo Likelihood Analysis}

We use CosmoMC in a Markov Chain Monte-Carlo
likelihood analysis \citep{Lewis:2002ah}. 
The parameter space that we explore spans the parameter set of
$\{H(0.35), D_A(0.35), \Omega_mh^2, \beta,\Omega_bh^2, n_s, \sigma_v,k_\star\}$. 
Only $\{H(0.35), D_A(0.35), \Omega_mh^2\}$ are well constrained using
SDSS LRGs alone. We marginalize over the other parameters, 
$\{\beta,\Omega_bh^2, n_s, \sigma_v,k_\star\}$, with the flat priors, 
$\{(0.1,0.6),(0.01859,0.02657),(0.865,1.059),(0,500)$km s$^{-1},(0.09,0.13)h$Mpc$^{-1}\}$, 
where the flat priors of $\Omega_b h^2$ and $n_s$ are centered on 
the measurements from WMAP7 and has width of $\pm7\sigma_{WMAP}$ (with $\sigma_{WMAP}$ from
\cite{Komatsu:2010fb}). These priors
are wide enough to ensure that CMB constraints are not double counted 
when our results are combined with CMB data \citep{Chuang:2010dv}.
We also marginalize over the amplitude of the galaxy correlation function, effectively
marginalizing over a linear galaxy bias.

\section{Results} \label{sec:results}

We now present the model independent measurements of the parameters
$\{H(0.35)$, $D_A(0.35)$, $\Omega_m h^2\}$, obtained by using the 
method described in previous sections. We also present the derived 
parameters including $H(0.35)\,r_s(z_d)/c$, $D_A(0.35)/r_s(z_d)$, 
$D_V(0.35)/r_s(z_d)$, and $A(0.35)$, where 
\begin{equation} \label{eq:dv}
 D_V(z)\equiv \left[(1+z)^2D_A^2\frac{cz}{H(z)}\right]^\frac{1}{3}
\end{equation} 
and
\begin{equation}\label{eq:A}
 A(z)\equiv D_V(z)\frac{\sqrt{\Omega_mH_0^2}}{cz}.
\end{equation}
We recommend using $\{H(0.35) \,r_s(z_d)/c, D_A(0.35)/r_s(z_d)\}$ instead of 
$\{H(0.35)$, $D_A(0.35)$, $\Omega_m h^2\}$ because they are more robust 
measurements from this study (see Sec. \ref{sec:test} for more detail).
We apply our method to the 2D 2PCF of the LasDamas 
mock catalogs and find that our measurements are consistent with the input 
parameters of the simulations. 

\begin{figure*}
\centering
\includegraphics[width=1 \linewidth,clip]
{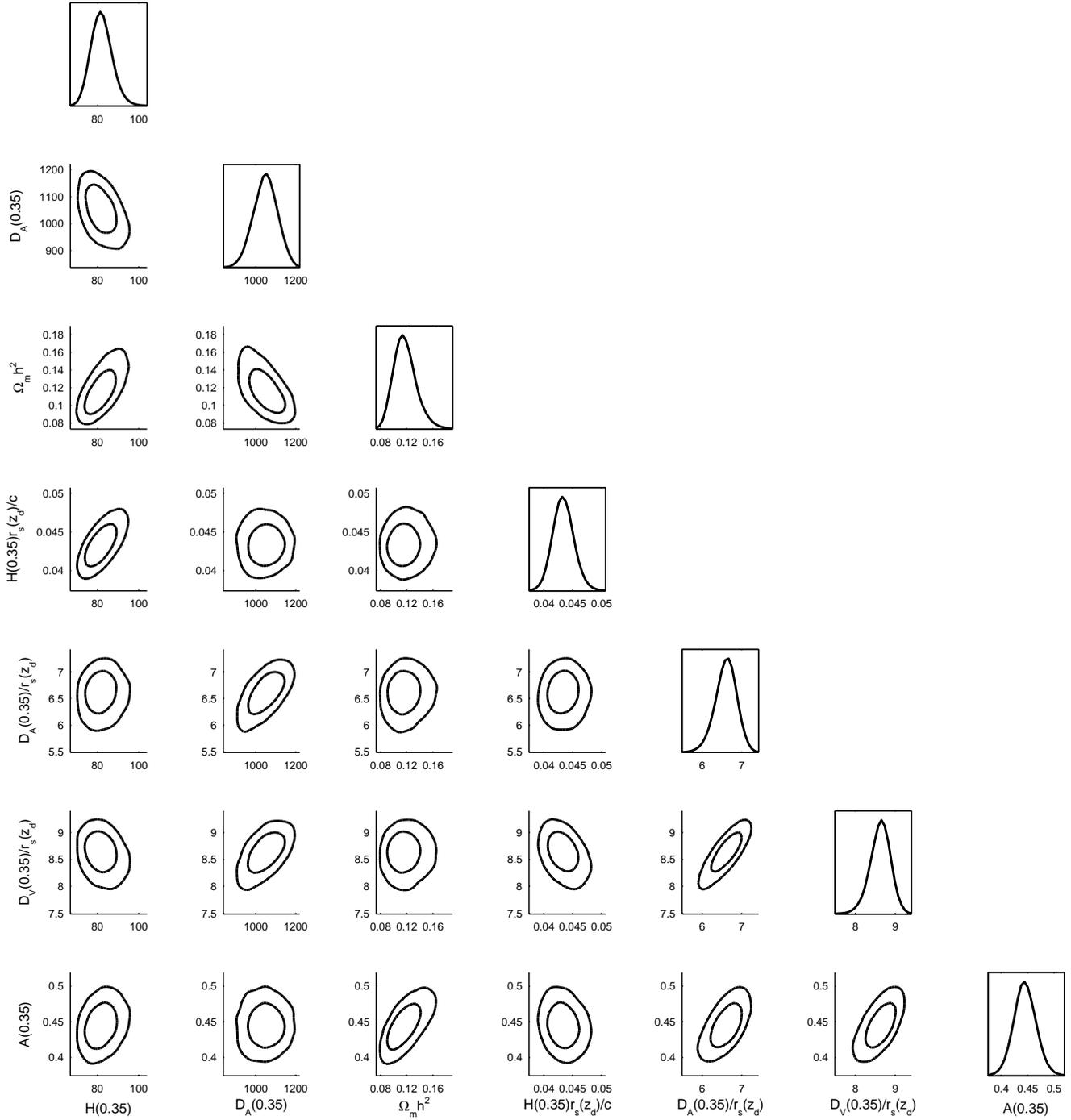}
\caption{2D marginalized
contours ($68\%$ and $95\%$ C.L.) for $\{H(0.35)$, $D_A(0.35)$, $\Omega_m h^2$, 
$H(0.35) \,r_s(z_d)/c$, $D_A(0.35)/r_s(z_d)$, $D_V(0.35)/r_s(z_d)$, $A(0.35)\}$. The diagonal
  panels represent the marginalized probabilities.
The unit of $H$ is $\Hunit$. The unit of $D_A$, $D_V$, and $r_s(z_d)$ is $\rm Mpc$.}
  \label{fig:sdss7params}
\end{figure*}

\begin{table}
\begin{center}
\begin{tabular}{crrrr}\hline
&mean &$\sigma$ &lower &upper \\ \hline
$	H(0.35)	$&\ \ 	82.1 &\ \  5.0 &\ \   77.2 &\ \   86.9	\\
$	D_A(0.35)$&\ \ 	1048  &\ \ 58 &\ \   990 &\ \  1107	\\
$	\Omega_m h^2	$ &\ \ 	0.118 &\ \  0.017 &\ \  0.101 &\ \  0.133	\\
\hline
$	H(0.35) \,r_s(z_d)/c	$ &\ \ 	{\bf 0.0434} &\ \  {\bf 0.0018} &\ \  {\bf 0.0417} &\ \  {\bf 0.0451}\\
$	D_A(0.35)/r_s(z_d)	$ &\ \ 	{\bf 6.60} &\ \  {\bf 0.26} &\ \  {\bf 6.34} &\ \  {\bf 6.85}\\
\hline
$	D_V(0.35)/r_s(z_d)	$ &\ \ 	8.62 &\ \  0.25 &\ \  8.38 &\ \  8.86\\
$	A(0.35)	$ &\ \ 	0.445	&\ \ 	0.021	&\ \ 	0.425	&\ \ 	0.465	\\
\hline
\end{tabular}
\end{center}
\caption{The mean, standard deviation, and the 68\% C.L. bounds of 
$\{H(0.35)$, $D_A(0.35)$, $\Omega_m h^2$, 
$H(0.35) \,r_s(z_d)/c$, $D_A(0.35)/r_s(z_d)$, $D_V(0.35)/r_s(z_d)$, $A(0.35)\}$ from SDSS DR7 LRGs. 
We recommend using $H(0.35) \,r_s(z_d)/c$ and $D_A(0.35)/r_s(z_d)$ for further analysis.
The unit of $H$ is $\Hunit$. The unit of $D_A$, $D_V$, and $r_s(z_d)$ is $\rm Mpc$.
} \label{table:mean}
\end{table}

\begin{table*}
\begin{center} 
\begin{tabular}{crrrrrrr}\hline
       &$H(0.35)$ &$D_A(0.35)$   &$\Omega_mh^2$ &$H(0.35) \,r_s(z_d)/c$ &$D_A(0.35)/r_s(z_d)$ &  $D_V(0.35)/r_s(z_d)$ &$A(0.35)$\\ \hline
$H(0.35)$		&\ \	1	&\ \ 	-0.4809	&\ \ 	0.7088	&\ \ 	0.7297	&\ \ 	0.0827	&\ \ 	-0.2631	&\ \ 	0.2618	\\
$D_A(0.35)$		&\ \	-0.4809	&\ \ 	1	&\ \ 	-0.6339	&\ \ 	-0.0065	&\ \ 	0.6730	&\ \ 	0.6167	&\ \ 	0.0379	\\
$\Omega_mh^2$		&\ \	0.7088	&\ \ 	-0.6339	&\ \ 	1	&\ \ 	0.0867	&\ \ 	0.0888	&\ \ 	0.0427	&\ \ 	0.7042	\\
$H(0.35) \,r_s(z_d)/c$ 	&\ \	0.7297 	&\ \   -0.0065  &\ \   0.0867  	&\ \  {\bf 1}  	&\ \ {\bf 0.0604} &\ \   -0.4104 	&\ \   -0.1934 	\\
$D_A(0.35)/r_s(z_d)$ 	&\ \	0.0827  &\ \   0.6730 	&\ \    0.0888 	&\ \ {\bf 0.0604}&\ \  {\bf 1} 	&\ \  0.8851  	&\ \   0.6447	\\
$D_V(0.35)/r_s(z_d)$	&\ \	-0.2631 &\ \   0.6167 	&\ \   0.0427   &\ \    -0.4104 &\ \   0.8851  	&\ \   1  	&\ \   0.6807	\\
$A(0.35)$		&\ \	0.2618  &\ \  0.0379  	&\ \   0.7042  	&\ \    -0.1934 &\ \   0.6447 	&\ \    0.6807 	&\ \    1	\\
\hline
\end{tabular}
\end{center}
\caption{Normalized covariance matrix of the measured and derived parameters, $\{H(0.35)$, $D_A(0.35)$, $\Omega_m h^2$, 
$H(0.35) \,r_s(z_d)/c$, $D_A(0.35)/r_s(z_d)$, $D_V(0.35)/r_s(z_d)$, $A(0.35)\}$. }
 \label{table:covar_matrix}
\end{table*}

\subsection{Constraints on $H(0.35)$ and $D_A(0.35)$ Independent of a Dark Energy Model} 
\label{sec:sdss_result}

Fig.\ \ref{fig:sdss7params} shows one and two-dimensional marginalized contours 
of the parameters, $\{H(0.35)$, $D_A(0.35)$, $\Omega_m h^2$, 
$H(0.35) \,r_s(z_d)/c$, $D_A(0.35)/r_s(z_d)$, $D_V(0.35)/r_s(z_d)$, $A(0.35)\}$,
derived in an MCMC likelihood analysis from the measured 2D 2PCF 
of the SDSS LRG sample.
Table \ref{table:mean} lists the mean, rms variance, and 68\%
confidence level limits of these parameters.
Table \ref{table:covar_matrix} gives the normalized covariance matrix
for this parameter set.
These are independent of a dark energy model, and obtained without
assuming a flat Universe.

The constraints on $\{H(0.35)$, $D_A(0.35)$, $\Omega_m h^2$, 
$H(0.35) \,r_s(z_d)/c$, $D_A(0.35)/r_s(z_d)$, $D_V(0.35)/r_s(z_d)$, $A(0.35)\}$, as summarized in
Tables \ref{table:mean} and \ref{table:covar_matrix},
can be used to combined with any other cosmological data set to
constrain dark energy and the cosmological model. We recommend using only
$\{H(0.35) \,r_s(z_d)/c, D_A(0.35)/r_s(z_d)\}$ since they have tighter constraints 
than $\{H(0.35),D_A(0.35)\}$ and are robust in the systematic tests we 
have carried out (see Sec. \ref{sec:test}).  
In addition, $H(0.35) \,r_s(z_d)/c$ and $D_A(0.35)/r_s(z_d)$ are basically independent to 
$\Omega_mh^2$ which might not a robust measurement in this study (see Sec.~\ref{sec:test}).

The bestfit model from the MCMC likelihood analysis has $\chi^2=112$
for 99 bins of data used 
for a set of 9 parameters             
(including the overall amplitude of the correlation function) and the $\chi^2$ per degree of freedom ($\chi^2/$d.o.f.) is 1.24.
Note that a $10\,h^{-1}$Mpc$\times10\,h^{-1}$Mpc bin is used if the center of the bin is in 
the scale range of $40 \,h^{-1}$Mpc$<s<120\,h^{-1}$Mpc.

\subsection{Validation Using Mock Catalogs} 
\label{sec:lasdamas_result}

In order to validate our method, we have applied it to 80 2D 2PCFs from
80 LasDamas mock catalogs (which are indexed with 01a-40a and 01b-40b). Again, 
we apply the flat and wide priors 
($\pm7\sigma_{WMAP7}$) on $\Omega_b h^2$ and $n_s$, centered on 
the input values of the simulation ($\Omega_b h^2=0.0196$ and $n_s=1$). 

\begin{figure}
\centering
\includegraphics[width=0.8 \columnwidth,clip,angle=270]{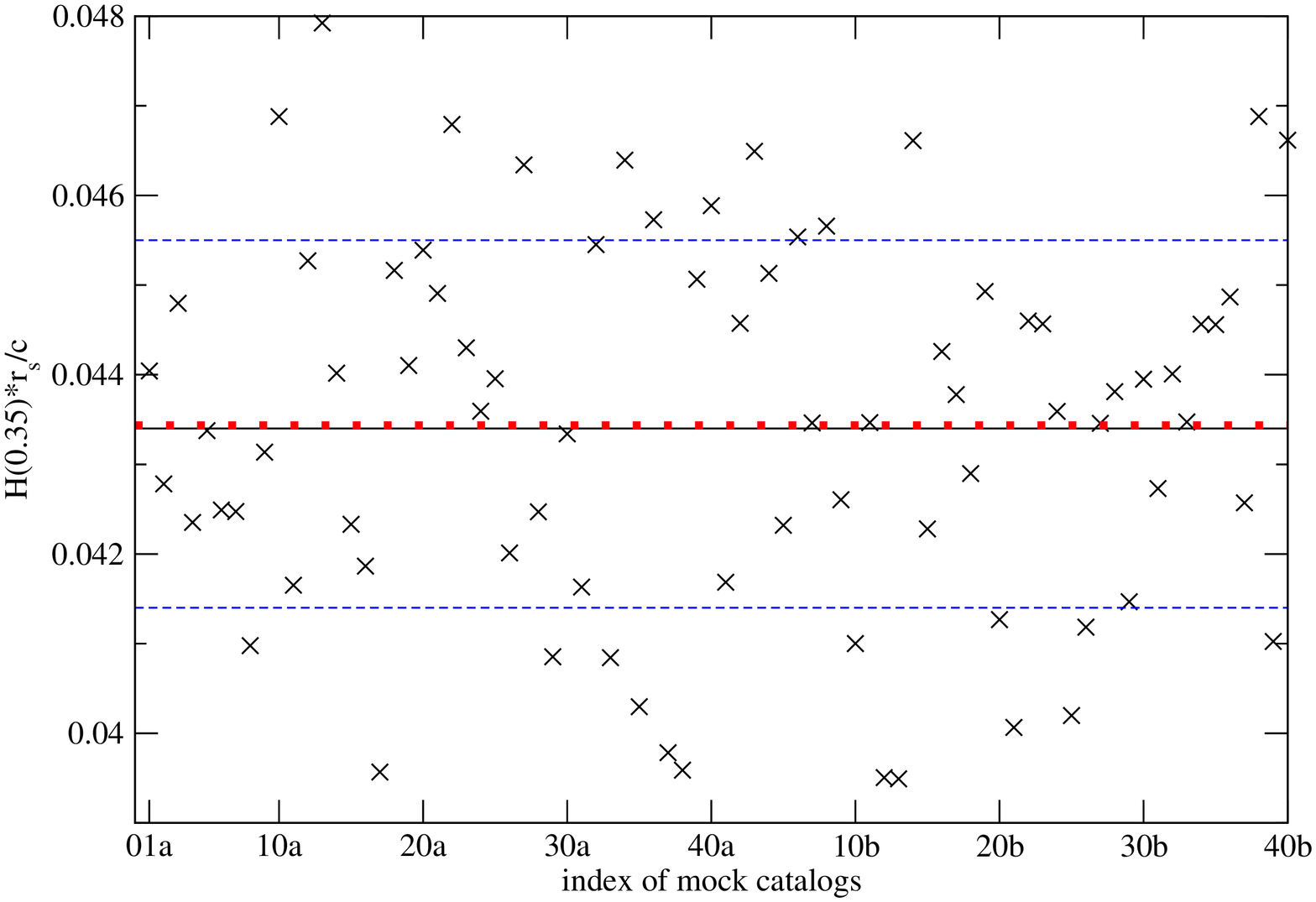}
\caption{Measurements of the means of $H(0.35) \,r_s(z_d)/c$ from 80 individual mock catalogs (indexed as 01a to 40a and 01b to 40b).
The black solid line shows the mean of these 80 measurements and the blue dashed lines show the range of $\pm \sigma$. The red dotted line shows the theoretical value computed with the input parameters of the simulations.
}
\label{fig:lasdamas_Hrs}
\end{figure}

\begin{figure}
\centering
\includegraphics[width=0.8 \columnwidth,clip,angle=270]{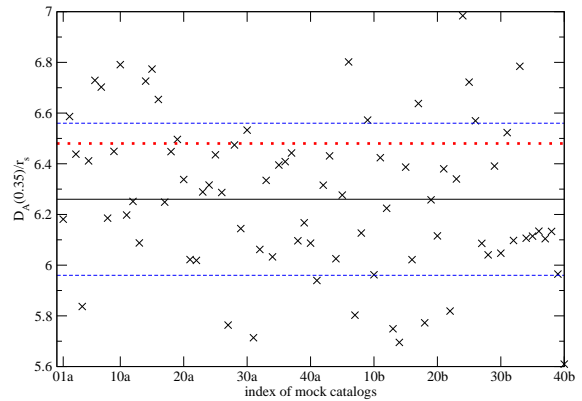}
\caption{Measurements of the means of $D_A(0.35)/r_s(z_d)$ from 80 individual mock catalogs (indexed as 01a to 40a and 01b to 40b).
The black solid line shows the mean of these 80 measurements and the blue dashed lines show the range of $\pm \sigma$. The red dotted line shows the theoretical value computed with the input parameters of the simulations.
}
\label{fig:lasdamas_rsbyDA}
\end{figure}

Table \ref{table:mean_lasdamas} shows the means and standard deviations of 
the distributions of our measurements of $\{H(0.35)$, $D_A(0.35)$, $\Omega_m h^2$, 
$H(0.35) \,r_s(z_d)/c$, $D_A(0.35)/r_s(z_d)$, $D_V(0.35)/r_s(z_d)$, $A(0.35)\}$ from 
each of the LasDamas mock catalogs (80 total) of the SDSS LRG sample. 
These are consistent with the input parameters, establishing the validity of our method. 
We also show the measurements 
of $H(0.35) \,r_s(z_d)/c$ and $D_A(0.35)/r_s(z_d)$ of each mock catalog in Fig.\ \ref{fig:lasdamas_Hrs} 
and Fig.\ \ref{fig:lasdamas_rsbyDA}. 
One can see the measurement of $H(0.35) \,r_s(z_d)/c$ and $D_A(0.35)/r_s(z_d)$ are consistent with the input parameters of the simulations.


\begin{table}
\begin{center}
\begin{tabular}{crr|r}\hline
&mean &$\sigma$ &input value\\ \hline
$H(0.35)$&\ \ 	81.1   &\ \     4.1    &\ \     81.79 \\
$D_A(0.35)$&\ \  1009   &\ \     56    &\ \    1032.8 \\
$\Omega_m h^2$&\ \ 0.121  &\ \     0.013 &\ \  0.1225\\
\hline
$H(0.35) \,r_s(z_d)/c$&\ \  0.0434  &\ \  0.0020   &\ \  0.0434\\
$D_A(0.35)/r_s(z_d)$&\ \ 6.26  &\ \ 0.30 &\ \ 6.48\\
\hline
$D_V(0.35)/r_s(z_d)$&\ \ 8.33   &\ \    0.31 &\ \  8.51\\
$A(0.35)$&\ \ 0.440   &\ \    0.019 &\ \  0.452\\
\hline
\end{tabular}
\end{center}
\caption{The mean, standard deviation, and the 68\% C.L. bounds of the distributions of
the measured values of $\{H(0.35)$, $D_A(0.35)$, $\Omega_m h^2$, 
$H(0.35) \,r_s(z_d)/c$, $D_A(0.35)/r_s(z_d)$, $D_V(0.35)/r_s(z_d)$, $A(0.35)\}$ from 
the 2D 2PCF of each of 80 LasDamas mock catalogs (which are indexed with 01a-40a and 01b-40b). 
Our measurements are consistent with the 
input values within 1$\sigma$, where each $\sigma$ is computed from the 80 means measured 
from the 80 mock catalogs. 
The unit of $H$ is $\Hunit$. The unit of $D_A$, $D_V$, and $r_s(z_d)$ is $\rm Mpc$.
} \label{table:mean_lasdamas}
\end{table}

\subsection{Cross-check with Measurements from Multipoles of the Correlation Function} 
\label{sec:result_multipoles}
As a cross-check of our results,
we now present the measurements from the monopole-quadrupole of the correlation function for comparason. The detail of the method is described in Appendix \ref{sec:multipoles}. Table\ \ref{table:mean_quad} lists the mean, rms variance, and 68\%
confidence level limits of these parameters. The measurements are consistent with those from our main method (see Table\ \ref{table:mean}). However, the constraints are much weaker ($> 8\%$, which is twice as large as our main results). 
This is most likely due to the fact that the information used in the monopole-quadrupole method is much less than what we use in our  main method (as presented in this paper). It is possible to obtain better measurements using multipole method by including higher order multipoles of the correlation function;
this is explored in \cite{Chuang:2012ad}.


\begin{table}
\begin{center}
\begin{tabular}{crrrr}\hline
&mean &$\sigma$ &lower &upper \\ \hline
$	H(0.35)	$&\ \ 	79.6	&\ \ 	8.8	&\ \ 	70.9	&\ \ 	87.8	\\
$	D_A(0.35)$&\ \ 	1060	&\ \ 	92	&\ \ 	970	&\ \ 	1150	\\
$	\Omega_m h^2	$&\ \ 	0.103	&\ \ 	0.015	&\ \ 	0.088	&\ \ 	0.118	\\
\hline
$	H(0.35) \,r_s(z_d)/c	$&\ \ 	0.0435	&\ \ 	0.0045	&\ \ 	0.0391	&\ \ 	0.0477	\\
$	D_A(0.35)/r_s(z_d)	$&\ \ 	6.44	&\ \ 	0.51	&\ \ 	5.99	&\ \ 	6.90	\\
\hline
\end{tabular}
\end{center}
\caption{The mean, standard deviation, and the 68\% C.L. bounds of 
$\{H(0.35)$, $D_A(0.35)$, $\Omega_m h^2$, 
$H(0.35) \,r_s(z_d)/c$, $D_A(0.35)/r_s(z_d)\}$ from SDSS DR7 LRGs using monopole-quadrupole of the correlation function. 
The bin size is 5 h$^{-1}$Mpc and the scale range is $40 < s < 120$ h$^{-1}$Mpc. $\chi^2/$d.o.f. is 1.23.
The unit of $H$ is $\Hunit$. The unit of $D_A$, $D_V$, and $r_s(z_d)$ is $\rm Mpc$.
} \label{table:mean_quad}
\end{table}

\section{SYSTEMATIC TESTS} 
\label{sec:test}

Table.\ \ref{table:test} shows the systematic tests that we have
carried out varying key assumptions made in our analysis.
These include the range of scales used to calculate the correlation function, 
the nonlinear damping factor, the bin size, and an overall shift in the measured correlation 
function due to a systematic error.

First, we fix the nonlinear damping factor, $k_\star=0.11h$Mpc$^{-1}$, and find the results 
are basically the same. To speed up the computation, we fix $k_\star$ for the rest 
of the tests.

In this study, we marginalize over $\beta$ with a wide flat prior (0.1 to 0.6) since 
our method is not sensitive to $\beta$. We test fixing the value of $\beta$ to 0.35, 
which is close to the measurement from previous work with similar data but using different 
method \citep{Cabre:2008sz}, and find that our measurements of $H(0.35) \,r_s(z_d)/c$ and 
$D_A(0.35)/r_s(z_d)$ change by less than 1\% compared to that of marginalizing over
$\beta$.

We vary the range of the scale and find that $H(0.35) \,r_s(z_d)/c$ and $D_A(0.35)/r_s(z_d)$ 
are insensitive to it. However, $\Omega_m h^2$ is sensitive to the minimum scale 
chosen which could imply that the scale dependent bias or redshift distortion is 
distorting larger scale than we have expected. Therefore, we do not recommend to 
use $\Omega_m h^2$ from this study. In the case of $s=40-130h^{-1}$Mpc, 
$D_A(0.35)/r_s(z_d)$ is different from the fiducial result with about 2$\sigma$, 
which is likely due to systematic errors responsible for the anomalously high
tail in the spherically-averaged correlation function (see, e.g., \cite{Chuang:2010dv})
on large scales.

We vary the bin size to 8$h^{-1}$Mpc$\times$8$h^{-1}$Mpc and find 
$\chi^2/$d.o.f.$=1.72$, which can be explained by the increase in the noise level 
with the increased number of bins.
The number of the mock catalogs used to construct the covariance matrix is 160 and 
the number of bins used with bin size = 8$h^{-1}$Mpc$\times$8$h^{-1}$Mpc is 159.
One can expect the covariance matrix would be too noisy to give reasonalbe results.

We also show that the results are insensitive to the constant shift by lowering down 
the data points of the observed correlation function by 0.001 and 0.002. 

\begin{table*}\scriptsize
\begin{center}
\begin{tabular}{cllllllll}
\hline
&$H(0.35)$ &$D_A(0.35)$  &$\Omega_m h^2$ &$\frac{H(0.35) \,r_s(z_d)}{c}$ &$\frac{D_A(0.35)}{r_s(z_d)}$ &$\frac{D_V(0.35)}{r_s(z_d)}$&$A(0.35)$&$\chi^2$/d.o.f.\\ \hline
fiducial result
&\ \ $	82.1	_{	-4.9	}^{+	4.8	}$
&\ \ $	1048	_{	-58	}^{+	59	}$
&\ \ $	0.118	\pm0.016$
&\ \ $	0.0434	\pm0.0017$
&\ \ $	6.60	\pm0.25$
&\ \ $	8.62	\pm0.24$
&\ \ $	0.445	\pm0.020$
&\ 1.24\\
						
$k_\star=0.11$						
&\ \ $	81.7	_{	-5.0	}^{+	4.9	}$
&\ \ $	1051	\pm59$
&\ \ $	0.116	_{	-0.017	}^{+	0.016	}$
&\ \ $	0.0434	\pm0.0017$
&\ \ $	6.59	\pm0.25$
&\ \ $	8.62	\pm0.24$
&\ \ $	0.443	\pm0.020$
&\ 1.24\\						
						
$\beta=0.35$						
&\ \ $	83.9	\pm5.6$
&\ \ $	1008	\pm53$
&\ \ $	0.132	\pm0.020$
&\ \ $	0.0430	_{	-0.0017	}^{+	0.0016	}$
&\ \ $	6.54	\pm0.24$
&\ \ $	8.60	\pm0.24$
&\ \ $	0.456	\pm0.021$
&\ 1.25\\						
						
$30<s<120, k_\star=0.11$						
&\ \ $	83.4	_{	-4.7	}^{+	4.5	}$
&\ \ $	1038	_{	-51	}^{+	52	}$
&\ \ $	0.120	_{	-0.014	}^{+	0.013	}$
&\ \ $	0.0437	_{	-0.0018	}^{+	0.0019	}$
&\ \ $	6.59	_{	-0.22	}^{+	0.23	}$
&\ \ $	8.59	_{	-0.23	}^{+	0.24	}$
&\ \ $	0.446	\pm0.016$
&\ 1.24\\						
						
$50<s<120, k_\star=0.11$						
&\ \ $	83.9	_{	-5.4	}^{+	5.5	}$
&\ \ $	1012	_{	-64	}^{+	63	}$
&\ \ $	0.134	_{	-0.023	}^{+	0.024	}$
&\ \ $	0.0428	\pm0.0019$
&\ \ $	6.59	\pm0.26$
&\ \ $	8.65	_{	-0.23	}^{+	0.24	}$
&\ \ $	0.460	\pm0.024$
&\ 1.06\\						
						
$40<s<110, k_\star=0.11$						
&\ \ $	80.6	\pm5.1$
&\ \ $	1087	_{	-60	}^{+	59	}$
&\ \ $	0.115	\pm0.016$
&\ \ $	0.0429	\pm0.0019$
&\ \ $	6.78	_{	-0.26	}^{+	0.27	}$
&\ \ $	8.81	\pm0.27$
&\ \ $	0.454	_{	-0.021	}^{+	0.022	}$
&\ 1.09\\						
						
$40<s<130, k_\star=0.11$						
&\ \ $	84.8	_{	-6.3	}^{+	6.4	}$
&\ \ $	987	_{	-60	}^{+	61	}$
&\ \ $	0.115	\pm0.016$
&\ \ $	0.0451	\pm0.0026$
&\ \ $	6.17	_{	-0.26	}^{+	0.27	}$
&\ \ $	8.14	\pm0.28$
&\ \ $	0.418	\pm0.019$
&\ 1.31\\						
						
bin size = 8$h^{-1}$Mpc$\times$8$h^{-1}$Mpc						
&\ \ $	87.9	_{	-6.0	}^{+	5.6	}$
&\ \ $	1037	\pm60$
&\ \ $	0.139	_{	-0.018	}^{+	0.017	}$
&\ \ $	0.0447	_{	-0.0024	}^{+	0.0023	}$
&\ \ $	6.78	\pm0.27$
&\ \ $	8.70	_{	-0.29	}^{+	0.27	}$
&\ \ $	0.470	_{	-0.020	}^{+	0.021	}$
&\ 1.72\\						
						
shift$=0.001, k_\star=0.11$						
&\ \ $	83.0	_{	-5.3	}^{+	5.4	}$
&\ \ $	1041	_{	-60	}^{+	61	}$
&\ \ $	0.124	_{	-0.018	}^{+	0.019	}$
&\ \ $	0.0433	_{	-0.0018	}^{+	0.0018	}$
&\ \ $	6.63	\pm0.26$
&\ \ $	8.65	_{	-0.26	}^{+	0.25	}$
&\ \ $	0.453	\pm0.021$
&\ 1.25\\						
						
shift$=0.002, k_\star=0.11$						
&\ \ $	85.2	_{	-5.6	}^{+	5.4	}$
&\ \ $	1024	_{	-65	}^{+	63	}$
&\ \ $	0.135	_{	-0.020	}^{+	0.021	}$
&\ \ $	0.0435	\pm0.0019$
&\ \ $	6.67	\pm0.28$
&\ \ $	8.67	\pm0.28$
&\ \ $	0.463	\pm0.022$
&\ 1.24\\
\hline
\end{tabular}
\end{center}
\caption{This table shows the systematic tests with  the damping factor, the scale range, the bin size,
and the assumed constant shift from a systematic error ($\xi_{obs}(s)=\xi_{true}(s)+$shift). The
fiducial results are obtained by considering the scale range ($40<s<120\ h^{-1}$Mpc), the bin size = 10$h^{-1}$Mpc$\times$10$h^{-1}$Mpc, and the damping factor,$k_{\star}$, marginalized over with the a flat prior ($0.09<k_\star<0.13\ h$Mpc$^{-1}$).
The other results are calculated with only specified quantities different from the fiducial one.
The unit of $H$ is $\Hunit$. The unit of $D_A$, $D_V$, and $r_s(z_d)$ is $\rm Mpc$. The unit of $k_\star$ is $h$Mpc$^{-1}$.
} \label{table:test}
\end{table*}

\section{Conclusion and Discussion} 
\label{sec:conclusion}

We have obtained the first measurements of $H(z)$ and $D_A(z)$ from galaxy 
clustering data in an MCMC likelihood analysis. Our constraints for 
the measured and derived parameters, $\{H(0.35)$, $D_A(0.35)$, $\Omega_m h^2$, 
$H(0.35) \,r_s(z_d)/c$, $D_A(0.35)/r_s(z_d)$, $D_V(0.35)/r_s(z_d)$, $A(0.35)\}$,
from the 2D 2PCF of the sample of SDSS DR7 LRGs  
are summarized by Tables\ \ref{table:mean} and \ref{table:covar_matrix}.
Our results are robust and independent of a dark energy model, and obtained without
assuming a flat Universe, and represent the first measurements of
$H(z)$ and $D_A(z)$ from galaxy clustering data. 

Our galaxy clustering measurements of $H(0.35) \,r_s(z_d)/c$ and $D_A(0.35)/r_s(z_d)$ (see 
Tables\ \ref{table:mean} and \ref{table:covar_matrix}) can be used to 
combine with CMB and other cosmological
data sets to probe dark energy. 
In a companion paper \citep{Wang:2011sb}, 
we explore the implications of our results for dark energy constraints.

We recommend using $H(0.35) \,r_s(z_d)/c$ and $D_A(0.35)/r_s(z_d)$ measured from 
the SDSS LRGs for combination with other data sets, since they are tight constraints 
(4\%) that
are nearly uncorrelated, and robust with respect to tests of systematic 
uncertainties.
This is as expected, 
since $H(0.35) \,r_s(z_d)/c$ and $D_A(0.35)/r_s(z_d)$
correspond to the preferential redshift separation along the line of sight,
and the preferential angular separation in the transverse direction respectively;
these in turn arise from the BAO in the radial and transverse directions.
The measurable preferential redshift and angular separations
should be uncorrelated since they are independent degrees of freedom.
On the other hand, the measurements of $H(0.35)$ and $D_A(0.35)$ are mainly determined by the geometrical distortion (i.e. the 2D correlation function is supposed to be isotropic without considering the redshift distortion), so that they are highly correlated (correlation coefficient $r\sim -0.5$).
The presence of the BAO (although only marginally visible in Fig.1)
leads to tight and robust constraints on $H(0.35) \,r_s(z_d)/c$ and $D_A(0.35)/r_s(z_d)$.
Since most of the constraining power in our analysis comes from fitting 
the overall shape of the galaxy correlation function on quasi-linear scales, 
and not from fitting the BAO peaks, we refer to our measurements as galaxy
clustering measurements.

We have validated our method by applying it to the 2D 2PCF of the mock 
catalogs from LasDamas, and finding consistency between our measurements 
and the input parameters of the LasDamas simulations for samples
(see Table \ref{table:mean_lasdamas}). 

As a cross-check of our results, we have measured $H(z)$ and $D_A(z)$ using monopole-quadrupole method and find that the results are consisent with our main method in this study but the constraints are much weaker. The reason is most likely that the information used 
by the monopole-quadrupole method is much less than what we use in our main method. However, it is still possible to improve the constraints by including higher order multipoles. 
We explore this issue in \cite{Chuang:2012ad}.

Our work has significant implications for future surveys in
establishing the feasibility of measuring both $H(z)$ and $D_A(z)$
from galaxy clustering data. 
In future work, we will optimize our method, and apply it to
new observational data as they become available, and to simulated
data of planned surveys to derive robust forecasts for dark energy
constraints.

\section*{Acknowledgements}
We would like to thank Chris Blake for useful comments.
We are grateful to the LasDamas project              
for making their mock catalogs publicly available.
The computing for this project was performed at the OU 
Supercomputing Center for Education and Research (OSCER) at the University of 
Oklahoma (OU). OSCER Director Henry Neeman and HPC Application Software Specialist 
Joshua Alexander provided invaluable technical support.  
This work was supported in part by DOE grant DE-FG02-04ER41305.

Funding for the Sloan Digital Sky Survey (SDSS) has been provided by the Alfred P. Sloan Foundation, the Participating Institutions, the National Aeronautics and Space Administration, the National Science Foundation, the U.S. Department of Energy, the Japanese Monbukagakusho, and the Max Planck Society. The SDSS Web site is http://www.sdss.org/.

The SDSS is managed by the Astrophysical Research Consortium (ARC) for the Participating Institutions. The Participating Institutions are The University of Chicago, Fermilab, the Institute for Advanced Study, the Japan Participation Group, The Johns Hopkins University, Los Alamos National Laboratory, the Max-Planck-Institute for Astronomy (MPIA), the Max-Planck-Institute for Astrophysics (MPA), New Mexico State University, University of Pittsburgh, Princeton University, the United States Naval Observatory, and the University of Washington. 
\setlength{\bibhang}{2.0em}

\appendix
\section{Algorithm of Smoothing the Covariance Matrix}
The original covariance matrix, $C_{ij}$, is computed by Eq. (\ref{eq:covmat}). Since the correlation function is in two dimension, we could re-label the covariance matrix as a function of the indexes of two bins, $C(\sigma_i,\pi_i,\sigma_j,\pi_j)$. The covariance matrix would be noisy since it is constructed with a small number of the mock catalogs comparing to the number of bins used. One might not obtain converging results while applying MCMC analysis on the observed data or an individual mock catalog, especially considering larger scale range or larger number of bins. We find that smoothing the covariance matrix could solve the problem. We have also checked that the smoothing procedure would not introduce bias by comparing the results from applying the smoothed and the original covariance matrix on the averaged correlation function from 160 mock catalogs. The concept of our smoothing procedure is that the new value of an element of the covariance matrix would be determined by its original value and the values of its neighbor elements. For example, to smooth an array, $f[n]$, we assign the new value at the index, $n'$, by $\tilde{f}[n']=(1-p)\cdot f[n']+p\cdot (f[n'-1]+f[n'+1])/2$ where $0\leq p\leq 1$ . The goal is to make the value be closer to the mean of the neighbors. Notice that while one of the neighbors is not available (i.e. $f[n']$ is the first or last element of the array), we let $f[n']$ be fixed since the mean of the neighbors is not available. The algorithm can be expressed by eq. (\ref{eq:smooth_general})
\begin{figure*}
\begin{eqnarray}\label{eq:smooth_general}
 \tilde{C}(\sigma_i,\pi_i,\sigma_j,\pi_j)= (1-p)\cdot C(\sigma_i,\pi_i,\sigma_j,\pi_j) + 
\frac{p}{m}\cdot \left[ \begin{array}{c} 
(C(\sigma_i+\Delta s,\pi_i,\sigma_j,\pi_j)+C(\sigma_i-\Delta s,\pi_i,\sigma_j,\pi_j))\mbox{, if both elements left are available}\\
(C(\sigma_i,\pi_i+\Delta s,\sigma_j,\pi_j)+C(\sigma_i,\pi_i-\Delta s,\sigma_j,\pi_j))\mbox{, if both elements left are available}\\
(C(\sigma_i,\pi_i,\sigma_j+\Delta s,\pi_j)+C(\sigma_i,\pi_i,\sigma_j-\Delta s,\pi_j))\mbox{, if both elements left are available}\\
(C(\sigma_i,\pi_i,\sigma_j,\pi_j+\Delta s)+C(\sigma_i,\pi_i,\sigma_j,\pi_j-\Delta s))\mbox{, if both elements left are available}
\end{array} \right],
\end{eqnarray}
\raggedright
where $m$ is the number of the neighbor elements used which should be 0, 2, 4, 6, or 8, and $\Delta s$ is the size of the bin in one direction. We use $p=0.01$ and $\Delta s=10$ Mpc/h in this study, but $p=0$ if $m=0$. And then, we iterate this procedure for 10 times.
\end{figure*}

While eq. (\ref{eq:smooth_general}) can be applied on most of the elements of the covariance matrix, there are some special cases as decrible below.\\
(1) Diagonal elements: these elements are only determined by the nearby diagonal elements by eq. (\ref{eq:smooth_diagonal}), since the diagonal elements are supposed to be larger than their off-diagonal neighbors and should not be smoothed using the latter.
\begin{figure*}
\begin{eqnarray}\label{eq:smooth_diagonal}
 \tilde{C}(\sigma_i,\pi_i,\sigma_i,\pi_i)&=& (1-p)\cdot C(\sigma_i,\pi_i,\sigma_i,\pi_i) \\ \nonumber
&+&\frac{p}{m}\cdot \left[ \begin{array}{c} 
(C(\sigma_i+\Delta s,\pi_i,\sigma_i+\Delta s,\pi_j)+C(\sigma_i-\Delta s,\pi_i,\sigma_i-\Delta s,\pi_i))\mbox{, if both elements left are available}\\
(C(\sigma_i,\pi_i+\Delta s,\sigma_i,\pi_i+\Delta s)+C(\sigma_i,\pi_i-\Delta s,\sigma_i,\pi_i-\Delta s))\mbox{, if both elements left are available}
\end{array} \right],
\end{eqnarray}
\end{figure*}\\
(2) First off-diagonal elements: these elements are only determined by the first off-diagonal elements nearby, i.e. ${C}(\sigma_i+\Delta s,\pi_i,\sigma_i,\pi_i)$ would be assigned a new value by eq. (\ref{eq:smooth_off-diagonal}), since a first off-diagonal element could be very different from its neighboring diagonal elements and should not be smoothed using the latter.  
\begin{figure*}
\begin{eqnarray}\label{eq:smooth_off-diagonal}
 \tilde{C}(\sigma_i+\Delta s,\pi_i,\sigma_i,\pi_i)&=& (1-p)\cdot C(\sigma_i+\Delta s,\pi_i,\sigma_i,\pi_i) \\ \nonumber
&+&\frac{p}{m}\cdot \left[ \begin{array}{c} 
(C(\sigma_i+2\Delta s,\pi_i,\sigma_i+\Delta s,\pi_j)+C(\sigma_i,\pi_i,\sigma_i-\Delta s,\pi_i))\mbox{, if both elements left are available}\\
(C(\sigma_i+\Delta s,\pi_i+\Delta s,\sigma_i,\pi_i+\Delta s)+C(\sigma_i+\Delta s,\pi_i-\Delta s,\sigma_i,\pi_i-\Delta s))\mbox{, if both elements left are available}
\end{array} \right],
\end{eqnarray}
\end{figure*}

\section{Measuring $H$ and $D_A$ with Multipoles of the Correlation Function} \label{sec:multipoles}
Using monopole-quadrupole of power spectrum to break the degeneracy of $H(z)$ and $D_A(z)$ is introduced by \cite{Padmanabhan:2008ag}. \cite{Kazin:2011xt} tested the method of monopole-quadrupole of correlation function with mock catalogs. The method used in \cite{Kazin:2011xt} is similar but not exactly the same as our method. We describe our method below. 

\begin{figure}
\centering
\includegraphics[width=0.8 \columnwidth,clip,angle=-90]{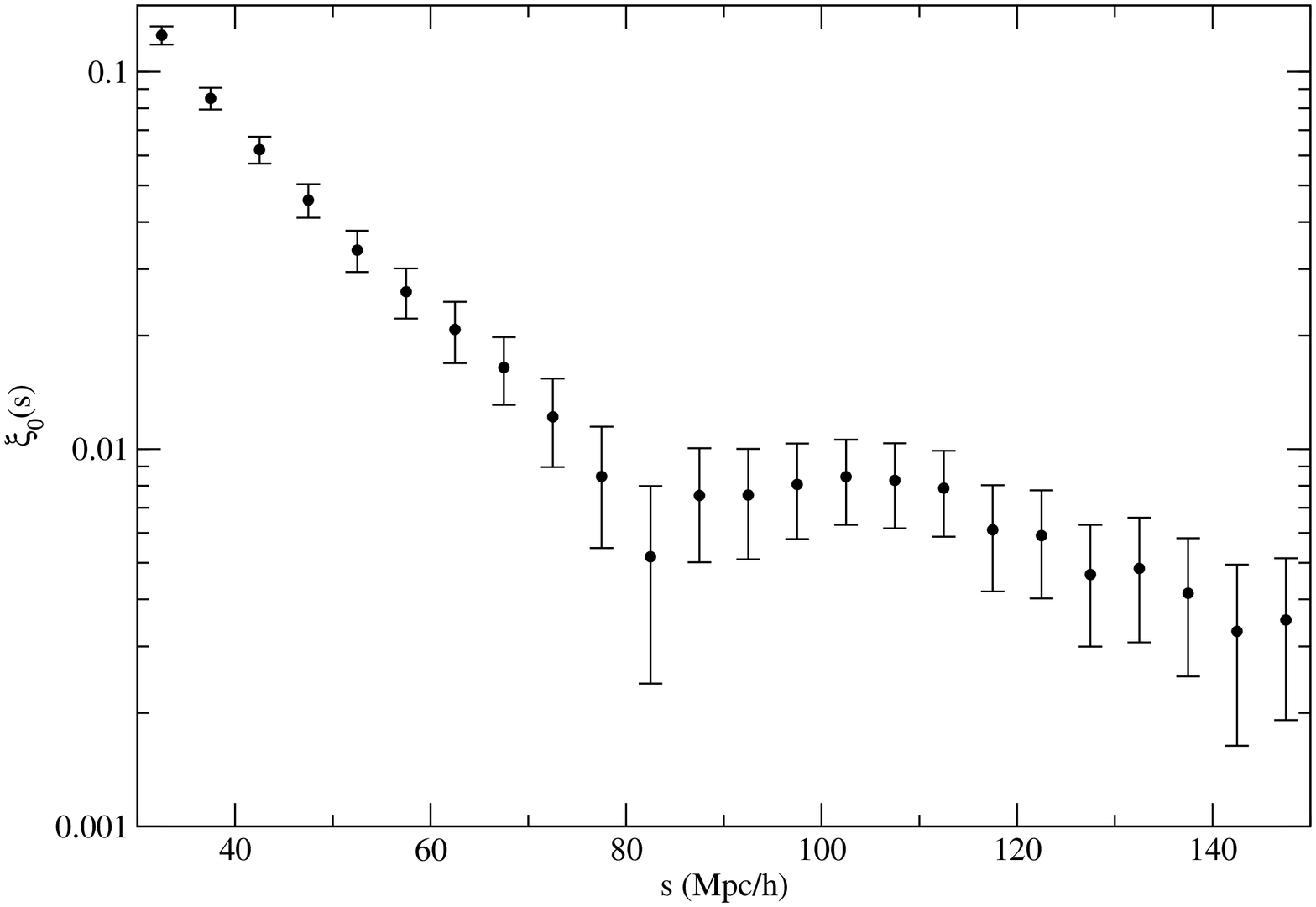}
\caption{Measurement of monopole of the correlation function from SDSS DR7 LRGs in a redshift range  $0.16 < z < 0.44$. The error bars are the square roots of the diagonal elements of the covariance matrix we have derived from mock catalogs.}
\label{fig:sdss_mono}
\end{figure}

\begin{figure}
\centering
\includegraphics[width=0.8 \columnwidth,clip,angle=-90]{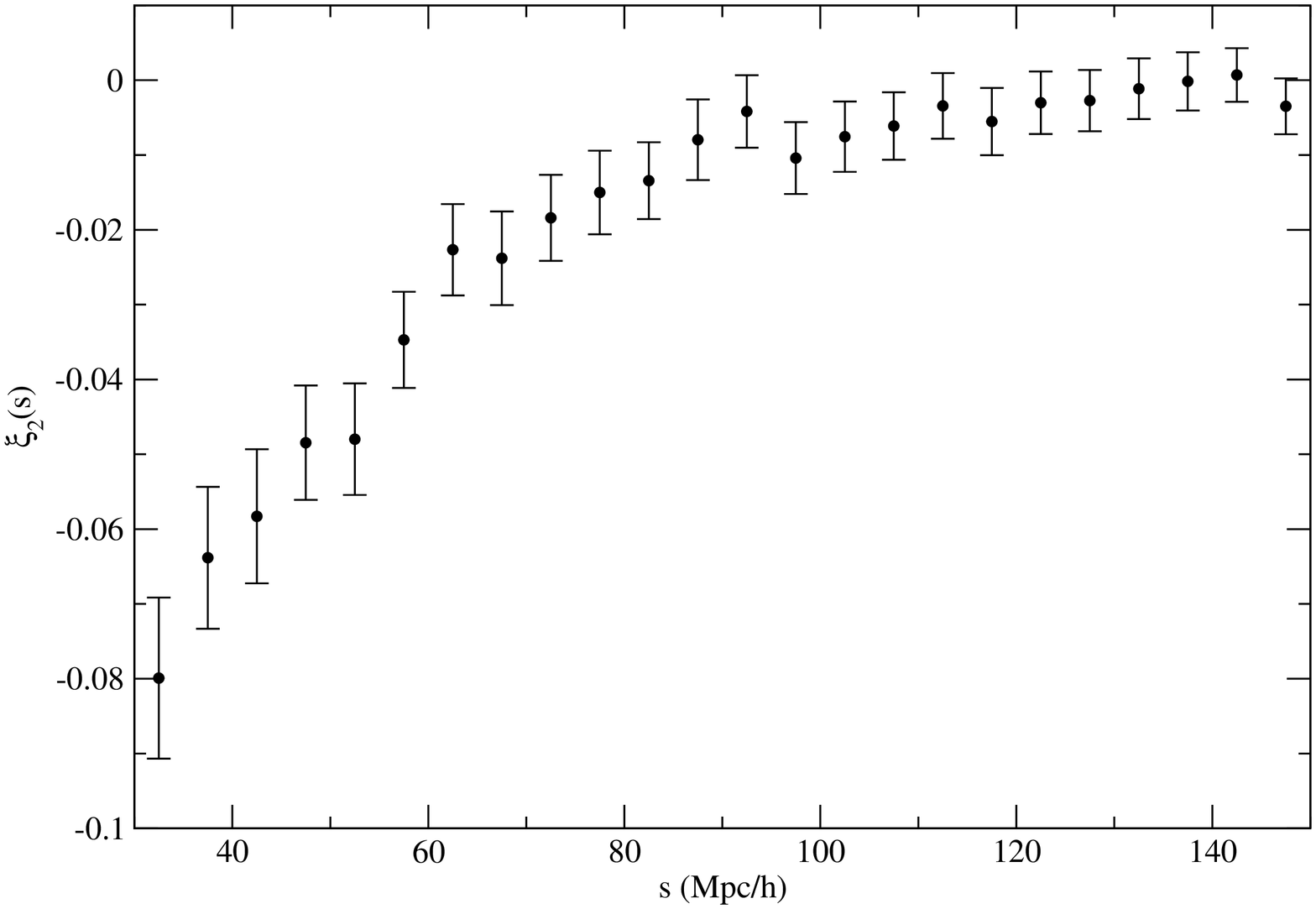}
\caption{Measurement of quadrupole of the correlation function from SDSS DR7 LRGs in a redshift range  $0.16 < z < 0.44$. The error bars are the square roots of the diagonal elements of the covariance matrix we have derived from mock catalogs.}
\label{fig:sdss_quad}
\end{figure}

First, we compute the 2D correlation function with bin size, $1 \, h^{-1}$Mpc$\times 1 \,h^{-1}$Mpc. Then, we compute the monopole and quadrupole by
\begin{equation}\label{eq:mono}
 \xi_0(s) \equiv \frac{\displaystyle\sum_{s-\frac{\Delta s}{2} < \sqrt{\sigma^2+\pi^2} < s+\frac{\Delta s}{2}}\xi(\sigma,\pi)\sqrt{1-\mu^2}}{\mbox{Number of bins used in the numerator}}
\end{equation}
and
\begin{equation}\label{eq:quad}
 \xi_2(s) \equiv \frac{\displaystyle\sum_{s-\frac{\Delta s}{2} < \sqrt{\sigma^2+\pi^2} < s+\frac{\Delta s}{2}}\frac{5}{2}\xi(\sigma,\pi)(3\mu^2-1)\sqrt{1-\mu^2}}{\mbox{Number of bins used in the numerator}},
\end{equation}
where 
\begin{equation}
\mu\equiv\frac{\pi}{\sqrt{\sigma^2+\pi^2}}
\end{equation}
and $\Delta s=5$ h$^{-1}$Mpc and the scale range used is $40 < s < 120$ h$^{-1}$Mpc in this study. Therefore, there are
32 data points for monopole-quadrupole method here, with 16 measurements each for monopole and quadrupole respectively.
Fig.\ \ref{fig:sdss_mono} and \ref{fig:sdss_quad} show the measurements of the monopole and quadrupole of the correlation function from the observed galaxy sample.

Just like our main method, the covariance matrix (for the 32 monopole and quadrupole measurements)
is constructed from the mock catalogs. The theoretical multipoles are computed 
using Eqs.\ (\ref{eq:mono}) and (\ref{eq:quad}).

Now, by exploring the same parameters and ranges using MCMC (with $\chi^2$ given by Eq.[\ref{eq:chi2_2}]), 
one could measure $H(z)$ and $D_A(z)$ following the same steps as our main method in this study. The results are shown in sec.\ \ref{sec:result_multipoles} (see Table 4). Like our main method, our monopole-quadrupole method has only one approximation, that the observed 2D correlation function using different fiducial models can be converted from one to another with two stretching factors. However, the multipole method tested in \cite{Kazin:2011xt} neglected some additional terms while measuring the stretching factors. Although the effect of the neglected terms might be small, it could be completely avoided by using the method described here.

\label{lastpage}

\end{document}